\documentstyle[prl,eqsecnum,aps]{revtex}

\begin{document}
\author{Jian-Qi Shen $^{1}$\footnote{E-mail address: jqshen@coer.zju.edu.cn} and Li-Hong Ma $^{2}$}
\address{$^{1}$ State Key Laboratory of Modern Optical Instrumentation, Center for Optical
and Electromagnetic Research; Zhejiang Institute of Modern Physics
and Department of Physics, Zhejiang University, Hangzhou 310027,
P. R. China
\\ $^{2}$ Institute of Information Optics and Department of Physics, Zhejiang Normal University, Zhejiang Jinhua 321004, P. R. China}
\date{\today }
\title{Geometric Phase and Helicity Inversion of Photons Propagating \\ inside a Noncoplanarly Curved Optical Fiber}
\maketitle

\begin{abstract}
The Letter presents an exact expression for the non-adiabatic
non-cyclic geometric phase of photons propagating inside a
noncoplanarly curved optical fiber by using the Lewis-Riesenfeld
invariant theory. It is shown that the helicity inversion of
photons arises in the curved fiber. Since we have exactly solved
the time-dependent Schr\"{o}dinger equation that governs the
propagation of photons in a curved fiber and, moreover, the
chronological product is not involved in this exact solution, our
formulation therefore has several advantages over other treatments
based on the classical Maxwell's theory and the Berry's adiabatic
quantum theory. The potential application of helicity inversion of
photons to information science is briefly suggested.



\end{abstract}

\section{Introduction}
In 1984, Berry found the existence of the Berry's topological
phase ( i. e., the adiabatic geometric phase ) of wave function in
adiabatic quantum process where the cyclic evolution of wave
function yields the original state plus a phase shift, which is a
sum of a dynamical phase and a geometric phase shift\cite {Berry}.
Berry's discovery opens up new opportunities for investigating the
global and topological properties of quantum evolution\cite
{Furtado,Shen1,Kuppermann,Wagh,Sanders}. It is now well known that
geometric phase arises in systems with the time-dependent
Hamiltonian, or in systems whose Hamiltonian possesses some
evolution parameters\cite{Shen2,Shen3}. Differing from dynamical
phase that depends on dynamical quantities of systems such as
energy, frequency, velocity as well as coupling coefficients,
geometric phase is independent of these dynamical quantities.
Instead, it is only related to the geometric nature of the pathway
along which quantum systems evolve. This, therefore, implies that
geometric phase presents the topological and global properties of
quantum systems in time-evolution process, and that it possesses
the physical significance and can thus be applied to various
fields of physics\cite{Gong,Taguchi,Falci}. Geometric phases
attract attentions of many physicists in considerable fields such
as gravity theory\cite {Furtado,Shen1}, differential
geometry\cite{Simon}, atomic and molecular physics\cite
{Kuppermann,Kuppermann2,Levi}, nuclear physics\cite {Wagh},
quantum optics\cite{Gong}, condensed matter
physics\cite{Taguchi,Falci}, molecular systems and chemical
reaction\cite {Kuppermann} as well.

The first physical realization of the Berry's geometric phase is
that of the polarized photon propagating inside the noncoplanar
optical fiber, of which whose theory is proposed by Chiao and
Wu\cite {Chiao}. Based on the experimental work of Chiao {\it et
al.}\cite{Tomita}, many researchers further studied this problem
in both experiments and
theories\cite{Kwiat,Robinson,Haldane1,Haldane2,ibid} afterwards,
where they tried to consider this geometric phase by using the
classical Maxwell$^{,}$s theory, differential geometry or quantum
mechanics. Although Chiao-Wu theory\cite {Chiao} concerning the
propagation of photons inside the optical fiber was very
successful in predicting this geometric phase, and in
satisfactorily investigating the polarization and propagation of
photons in the cyclic adiabatic process in the fiber experiment
performed by Kwiat and Chiao {\it et al.}\cite{Chiao,Kwiat}, we
still argue that there exist at least three points which should be
further discussed. They are given as follows:

(i) since it is based on Berry$^{,}$s formula of adiabatic
geometric phase\cite {Berry}, Chiao-Wu formulation can be
applicable only to the adiabatic and cyclic quantum process, i.
e., this method is not appropriate to deal with the general
non-adiabatic non-cyclic geometric quantum phase;

(ii) Chiao-Wu theory is first-quantized. However, we suggest the
second-quantized treatment of the photon in the fiber, so that
this second-quantized formulation enables us to consider the
geometric phase in the fiber at quantum level, and thus provides
us with a theoretical tool to deal with the hot arguments as to
whether this geometric phase inside the fiber belongs to the
classical or the quantum level\cite{Kwiat};

(iii) Chiao-Wu theory has no expression for the Hamiltonian
describing the interaction of photons field with fiber medium.
Instead, it investigated the geometric phase merely by taking into
consideration the eigenvalue equation of the helicity of photons
in curved fiber. Note, however, that this eigenvalue equation
cannot always be regarded as the Schr\"{o}dinger equation
governing the propagation of photons in the noncoplanarly curved
fiber, although in the sense of adiabatic process, it can be
considered the Schr\"{o}dinger equation indeed. This, therefore,
means that the Chiao-Wu formulation should be generalized to a
more complete one that can satisfactorily study the non-adiabatic
evolution of wave function of the photon in curved fiber.

In the present paper, we resolve these problems by making use of
the Lewis-Riesenfeld invariant theory and the invariant-related
unitary transformation formulation\cite{Riesenfeld,Gao1}. By
constructing an effective second-quantized Hamiltonian describing
the interaction between photons field and fiber medium, we obtain
the exact solution of the time-dependent Schr\"{o}dinger equation
and, based on this result, we obtain the non-adiabatic non-cyclic
geometric phase of photons, rather than the phase in the sense of
Berry's adiabatic quantum process. Note that the exact solution
presented here does not contain the chronological product
operator, namely, it may be considered the explicit solution. On
the basis of the above work, we consider the helicity inversion of
photons in the optical fiber, to which most of the methods based
on the Maxwell's theory, differential geometry ( parallel
transport )\cite{Simon} or Berry's phase theory\cite {Berry} is
not applicable since these methods are not in connection with the
time-evolution equation ( i. e., the time-dependent
Schr\"{o}dinger equation) of the photon in the fiber.

Time-dependent system is governed by the time-dependent
Schr\"{o}dinger equation. The invariant theory\cite{Gao1}
suggested by Lewis and Riesenfeld can solve the time-dependent
Schr\"{o}dinger equation. Gao {\it et al.} proposed a generalized
invariant theory\cite{Gao1,Gao2}, by introducing the basic
invariants which enable one to find the complete set of commuting
invariants for some time-dependent multi-dimensional
systems\cite{Gao2}. Since the exact solution to quantum systems
with time-dependent Hamiltonian obtained by the invariant-related
unitary transformation formulation contains both the geometric
phase and the dynamical phase and, fortunately, all these results
are explicit rather than formal ( i. e., there exists no
chronological product operator in wave function ), the
Lewis-Riesenfeld theory
thus developed into a powerful tool for treating the time-dependent Schr\"{o}%
dinger equation and the geometric phase factor.

\section{Time evolution and noncyclic geometric phase of photons in a curved fiber }
Consider a noncoplanarly curved optical fiber that is wound
smoothly on a large enough diameter\cite{Tomita}, an equation of
motion of the photon
propagating along the fiber is of the form

\begin{equation}
{\dot{\bf{k}}}+{\bf{k}}\times (%
\frac{{\bf{k}}\times \dot{\bf{k}}}{k^{2}})=0 \eqnum{1}
\label{eq100}
\end{equation}
with ${\bf{k}}(t)$ being the wave vector of the photon,
${\bf{k}}\times (\frac{\bf{k}\times \dot{\bf{k}}}{k^{2}})$ and
$\frac{ \bf{k}\times \dot{\bf{k}}}{k^{2}}$ may be considered the
generalized Lorentz magnetic force ( Coriolis force ) and the
generalized magnetic field strength, respectively. The reason for
${\bf{k}}\times (\frac{\bf{k}\times \dot{\bf{k}}}{k^{2}})$ being
called generalized Lorentz force is that Eq. (\ref{eq100}) is
somewhat analogous to the equation of motion of a photon moving in
the gravitomagnetic fields\cite{Qi}. Dot in the equation of motion
(\ref{eq100}) denotes the time rate of change of ${\bf{k}}(t)$.
Further analysis shows that the infinitesimal rotation operator of
motion of a photon in the fiber is given as
$U_{R}=1-i\bf{\vartheta}\cdot \bf{J}$ with $\bf{J}$ being the
total angular momentum operator of photons field and
${\bf{\vartheta}}=\frac{{\bf{k}}(t)\times \dot
{\bf{k}}(t)}{k^{2}}{\rm d}t$. It follows from the evolution
operator, $U_{R}$, that the photon state $\left| \sigma
,{\bf{k}}(t)\right\rangle $ $( \sigma $ denotes the eigenvalue of
$I(t)=\frac{{\bf{k}}(t)}{k}\cdot \bf{J}$ that is proved to be an
invariant in what follows )
satisfies the following time-evolution equation ( in the unit $%
c=\hbar =1$ )

\begin{equation}
i\frac{\partial \left| \sigma ,{\bf{k}}(t)\right\rangle }{\partial t}=\frac{%
{\bf{k}}(t)\times \dot{\bf{k}}(t)}{k^{2}}\cdot {\bf{J}}\left|
\sigma ,{\bf{k}}(t)\right\rangle        \eqnum{2}    \label{eq1}
\end{equation}%
with $H_{eff}(t)=\frac{{\bf{k}}(t)\times \dot{\bf{k}}(t)}{k^{2}}%
\cdot \bf{J}$ \ being the effective Hamiltonian of photon inside
the fiber. Apparently, $I(t)$ and the effective Hamiltonian
$H_{eff}(t)$ agree with the invariant equation ( Liouville-Von
Neumann equation )\cite{Riesenfeld}

\begin{equation}
\frac{\partial I(t)}{\partial t}+\frac{1}{i}[I(t),H_{eff}(t)]=0.
\eqnum{3} \label{eq202}
\end{equation}
Substitution of
the expressions for
 $I(t)$ and $H_{eff}(t)$ into this invariant equation yields the
 equation, ${\dot{\bf{k}}}+{\bf{k}}\times (\frac{%
\bf{k}\times \dot{\bf{k}}}{k^{2}})=0$, of motion of the photon in
the fiber. It follows that the eigenvalue of
$\frac{{\bf{k}}(t)}{k}\cdot \bf{J}$ of the photon is conserved in
motion and it is therefore an invariant in terms of the invariant
equation(\ref{eq202}). It can be seen from the form of
$H_{eff}(t)$ that the problem of the rotation of polarization
plane is actually in analogy with that of the time-dependent
quantum spin model\cite{Shen3}. Set the components of momentum of
a photon

\begin{equation}
\frac{{\bf{k}}(t)}{k}=(\sin \lambda (t)\cos \gamma (t),\sin
\lambda(t)\sin \gamma (t),\cos \lambda (t)),\eqnum{4}
\end{equation}
where the time-dependent parameters $ \lambda (t)$ and $\gamma
(t)$ denote the angle displacements of ${\bf{k}}(t)$
in the spherical polar coordinate system, and the invariant $I(t)=\frac{%
{\bf{k}}(t)}{k}\cdot \bf{J}$ may therefore be rewritten
\begin{equation}
I(t)=\frac{1}{2}\sin \lambda (t)\exp [-i\gamma
(t)]J_{+}+\frac{1}{2}\sin \lambda (t)\exp [i\gamma (t)]J_{-}+\cos
\lambda (t)J_{3}          \eqnum{5}
\end{equation}
with $J_{\pm }=J_{1}\pm iJ_{2},[J_{3},J_{\pm }]=\pm J_{\pm
},[J_{+},J_{-}]=2J_{3}$.

In order to obtain the analytical solution of the time-dependent
Schr\"{o}dinger equation (\ref{eq1}), we introduce an
invariant-related unitary transformation operator $V(t)$ that is
of the form

\begin{equation}
V(t)=\exp [\beta (t)J_{+}-\beta ^{\ast }(t)J_{-}], \eqnum{6}
\label{eq2}
\end{equation}
where the time-dependent parameters $\beta (t)=-\frac{\lambda (t)}{2}\exp [-i\gamma (t)],\quad \beta ^{\ast }(t)=-\frac{%
\lambda (t)}{2}\exp [i\gamma (t)]$. $V(t)$ can be easily shown to
transform the {\it time-dependent} invariant $I(t)$ to $I_{V}  \
(\equiv V^{\dagger }(t)I(t)V(t)=J_{3})$, which is {\it
time-independent}. The eigenstate of $I_{V}=J_{3}$ corresponding
to the eigenvalue $\sigma $ is denoted by $\left| \sigma
\right\rangle $. By making use of $V(t)$ in expression (\ref{eq2})
and the Baker-Campbell-Hausdorff formula\cite{Wei}, one can obtain
$H_{V}(t)$ from $H(t)$, i. e.,

\begin{eqnarray}
H_{V}(t) &=&V^{\dagger }(t)H(t)V(t)-V^{\dagger }(t)i\frac{\partial V(t)}{%
\partial t}  \nonumber \\
&=&\left\{{[\cos \lambda \cos \theta +\sin \lambda \sin \theta
\cos (\gamma -\varphi )]+\dot{\gamma}(1-\cos \lambda
)}\right\}J_{3}, \eqnum{3}  \eqnum{7}
 \label{eq3}
\end{eqnarray}%
where the time-dependent parameters $\theta $ and $\varphi $
represent the
angle displacements of $\frac{{\bf{k}}(t)\times \dot{\bf{k}}(t)}{%
k^{2}}$ in the spherical polar coordinate system, i. e.,
$\frac{{\bf{k}}(t)\times \dot{\bf{k}}(t)}{k^{2}}=(\sin \theta \cos
\varphi ,\sin \theta \sin \varphi ,\cos \theta ).$ It follows from
Eq. (\ref{eq3}) and the expression $I_{V}=J_{3}$ that $H_{V}(t)$
differs from $I_{V}$ only by a time-dependent $c$- number factor
$\exp [\frac{1}{i}\int_{0}^{t}\left\langle \sigma \right|
H_{V}(t^{\prime })\left| \sigma \right\rangle {\rm d}t^{\prime
}]$. By using the invariant equation (\ref{eq202}), the two
auxiliary equations, which are used to determine the
time-dependent parameters $\theta$ and $\varphi$, can be derived

\begin{equation}
\dot{\gamma}\sin ^{2}\lambda =\cos \theta ,\quad \dot{\lambda}\cos \gamma -\dot{%
\gamma}\cos \lambda \sin \lambda \sin \gamma =\sin \theta \sin
\varphi .        \eqnum{8}
\end{equation}
Fortunately, it is shown from the
auxiliary equations that

\begin{equation}
\cos \lambda \cos \theta +\sin \lambda \sin \theta \cos (\gamma
-\varphi )=0,     \eqnum{9}
\end{equation}
the expression (\ref{eq3}) for
$H_{V}(t)$ can thus be simplified into
$H_{V}(t)=\dot{\gamma}(t)[1-\cos \lambda (t)]J_{3}$.

Based on the invariant theory, the geometric phase of a photon
whose initial eigenvalue of helicity is $\sigma $ can be expressed
by

\begin{equation}
\phi _{\sigma }(t)=\left\{{\int_{0}^{t}\dot{\gamma}(t^{^{\prime
}})[1-\cos \lambda (t^{^{\prime }})]{\rm d}t^{^{\prime
}}}\right\}\left\langle \sigma \right| J_{3}\left| \sigma
\right\rangle . \eqnum{10}                  \label{eq4}
\end{equation}
Since it is easy to obtain the eigenvalues and eigenstates of
$I_{V}(t)=J_{3},$ with the help of (\ref{eq2}) and (\ref{eq3}),
one can arrive at the general solution of the time-dependent
Schr\"{o}dinger equation, which governs the motion of photons in
the fiber experiment. The result is given as follows

\begin{equation}
\left| \Psi (t)\right\rangle _{s}=\sum_{\sigma }C_{\sigma }\exp
[\frac{1}{i}\phi _{\sigma }(t)]V(t)\left| \sigma \right\rangle
\eqnum{11} \label{eq5}
\end{equation}
with the coefficients $C_{\sigma }=\langle \sigma ,t=0\left| \Psi
(0)\right\rangle _{s}.$ It is noted that the exact solution
presented here does not contain the chronological product and may
be considered the explicit solution.

In the noncoplanar optical fiber, according to the expression for
$H_{V}(t)$, the dynamical phase
\begin{equation}
\phi _{\sigma }^{({\rm d})}(t)=\left\{{\int_{0}^{t}\left[{\cos
\lambda(t') \cos \theta(t') +\sin \lambda(t') \sin \theta(t') \cos
(\gamma(t') -\varphi(t'))}\right]{\rm d}t^{^{\prime
}}}\right\}\left\langle \sigma \right| J_{3}\left| \sigma
\right\rangle       \eqnum{12}
\end{equation}
of the photon due to the effective Hamiltonian vanishes ( But note
that the Hamiltonian of free photons, $H_{0}=\frac{V}{(2\pi
)^{3}}\int {\rm d}{\bf k}\omega _{k}(a_{R}({\bf k})^{\dagger
}a_{R}({\bf k})+a_{L}({\bf k})^{\dagger }a_{L}({\bf k})+1)$ with
$a_{R}^{\dagger }(a_{R})$ and $a_{L}^{\dagger }(a_{L})$ being
respectively the creation (annihilation) operators for the right-
and left- polarized photons, which is not involved in the
effective Hamiltonian, may contribute a dynamical phase to the
wave function of photons in the curved fiber ), and its geometric
phase is expressed by the expression (\ref{eq4}). The time
evolution operator of wave function is $U(t)=\exp [\frac{1}{i}\phi
_{\sigma }(t)]V(t)$. For the sake of $U(t=0)=1$ and consequently
$\lambda (t=0)=0$, the initial conditions are taken to be
$k_{1}=0,k_{2}=0,k_{3}=k$. Thus the expectation value of the third
component of the orbital angular momentum $\bf{L}$ of the linear
polarized photon vanishes, namely, $\left\langle \sigma \right|
L_{3}\left| \sigma \right\rangle =\frac{{\bf{k}}(t)}{k}\cdot
\left\langle \sigma \right| {\bf{L}}\left| \sigma \right\rangle
=0$. This means that we should only analyze the expectation value
of the third component of spin of the photon. It follows from the
spin operator of non-normal product

\begin{eqnarray}
 {\bf{S}}&=&\frac{1}{2}\int {\rm
d}{\bf{k}}\frac{{\bf{k}}}{k}\left\{{[a_{R}({\bf k})^{\dagger
}a_{R}({\bf k})+a_{R}({\bf k})a_{R}({\bf k})^{\dagger
}]-[a_{L}({\bf k})^{\dagger }a_{L}({\bf k})+a_{L}({\bf
k})a_{L}({\bf k})^{\dagger
}]}\right\}       \nonumber \\
 &=& \int {\rm
d}{\bf{k}}\frac{{\bf{k}}}{k}\left\{{[a_{R}({\bf k})^{\dagger
}a_{R}({\bf k})+\frac{1}{2}]-[a_{L}({\bf k})^{\dagger }a_{L}({\bf
k})+\frac{1}{2}]}\right\} \eqnum{13}
\end{eqnarray}
that the total spin of the photons field comprises the
contributions of both the right- and left- rotation photons. It
can be found that the zero-point energy of both the right- and
left- rotation photons exists in the effective Hamiltonian. This
time-dependent zero-point energy possesses the physical meaning
and also contributes to the geometric phase factor. Thus, it
follows that there exists the quantum-vacuum geometric phase

\begin{equation}
\phi _{\pm}^{({\rm g})}=\pm
\frac{1}{2}\int_{0}^{t}\dot{\gamma}(t^{^{\prime }})[1-\cos \lambda
(t^{^{\prime }})]{\rm d}t^{^{\prime }} \eqnum{14}
\end{equation}
 ( $\pm$ corresponding to
the right- and left- polarized photon, respectively), in addition
to the classical geometric phase $\phi _{\pm}^{(c)}=\pm
\int_{0}^{t}\dot{\gamma}(t^{\prime })[1-\cos \lambda (t^{\prime
})]{\rm d}t^{^{\prime }}$ that has been measured in the fiber
experiment performed by Tomita and Chiao. However, it should be
pointed out that, even at the quantum level, the observable
quantum-vacuum geometric phase is absent in the fiber experiment,
since the sign of the quantum-vacuum geometric phase of the left-
and right- rotation photons is opposite, and their quantum-vacuum
geometric phase are therefore counteracted, hence the observable
geometric phase is merely the classical geometric phase, which,
arising as the adiabatic cyclic Berry's phase in the fiber
experiment, has been proved existing by Tomita and
Chiao\cite{Tomita}.

In the adiabatic and cyclic case, it is of physical interest to
discuss the topological property of the geometric phase arising in
the fiber experiment. Note that in the adiabatic and cyclic case
the expressions for geometric phase presented in this Letter is
reduced to that in the Chiao-Wu theory\cite{Chiao}, namely, in
this case our calculation is consistent with the result obtained
by Chiao and Wu. The adiabatic geometric phase ( Berry's phase )
in a cycle associated with $\gamma (t)$ in the parameter space of
the invariant $\frac{{\bf{k}}(t)}{k}\cdot \bf{J}$  is $\phi
_{\sigma }(T)=2\pi \sigma (1-\cos \lambda )$, where $\lambda $ is
taken to be time-independent just as what has been done in the
Chiao-Wu theory and the subsequent experiment performed by Tomita
{\it et al.}\cite{Tomita}. It is apparent that in the adiabatic
and cyclic geometric phase, $2\pi (1-\cos \lambda )$ is the solid
angle over the parameter space of the invariant, which are said to
represent the geometric meanings of the phase. This, therefore,
implies that geometric phase reflects the global and topological
properties of time-dependent evolution of quantum systems. It is
worthwhile to point out that although the adiabatic solution
describing the time evolution of photons in the curved fiber has
sufficient features to indicate the global and topological
properties of geometric phase of photons, the exact solution
obtained via the Lewis-Riesenfeld invariant theory possesses more
rich characters and can therefore treat more general
time-dependent evolution of wavefunctions in Quantum Mechanics
regimes in addition to the narrower class of the quantum adiabatic
process.

\section{Helicity inversion of photons in a curved fiber}
In the optical fiber that is wound smoothly on a large enough
diameter, the variation of the helicity of the photon cannot be
easily observed. However, in the sharply curved fiber, it is
apparent that the helicity depends strongly upon the geometric
shape of the fiber. In order to see whether or not this is the
true case, in what follows we deal with the helicity inversion of
photons inside the curved fiber. The principal reasons for the
importance of helicity inversion lie in that: (i) it is believed
that neither the adiabatic quantum theory nor the classical
electrodynamics can satisfactorily deal with this problem. Only
the explicit wave function of photons, which does not involve the
chronological product, in the optical fiber is obtained can we
deal with the time evolution of wave function and the helicity
inversion of photons moving along the curved fiber; (ii) helicity
inversion may be employed to the field of information science,
since the left-polarized laser ( representing $1$ ) may be
inverted to the right-polarized laser ( representing $0$ ) by the
mirror reflection and the optical helix. The Hamiltonian of free
photons field, ${\bf {p}}\cdot {\bf {v}}(t)$ with ${{\bf
{v}}(t)}=(\sin \lambda (t)\cos \gamma (t),\sin \lambda (t)\sin
\gamma (t),\cos \lambda (t))$, possesses the same eigenvalue of
$\bf {k}\cdot \bf {J}$ \ divided by $\sigma $. The particular
solution of Eq. (\ref{eq1}) corresponding to the eigenvalue,
$\sigma $ , is

\begin{equation}
\left| \sigma ,{\bf {k}}(t)\right\rangle  =V(t)\exp
\left[{\frac{1}{i}\int_{0}^{t}H_{V}(t^{\prime }){\rm d}t^{^{\prime
}}}\right]\left| \sigma \right\rangle \eqnum{15}
\end{equation}
that can be rewritten in the interaction representation as

\begin{equation}
\left| \sigma ,{\bf {k}}(t)\right\rangle =V(t)\exp
\left\{{\frac{1}{i}\int_{0}^{t}[H_{V}(t^{\prime })-{\bf {
k}}(t^{\prime })\cdot {\bf {J}}]{\rm d}t^{\prime }}\right\}\exp
\left\{{\frac{1}{2}\int_{0}^{t}{\rm d}t^{\prime }\int_{0}^{t}{\rm
d}t''[H_{V}(t^{\prime }),{\bf { k}}(t'')\cdot {\bf
{J}}]}\right\}\left| \sigma ,t\right\rangle_{I}   \eqnum{16}
\end{equation}
with $\left| \sigma ,t\right\rangle_{I} =\exp
[\frac{1}{i}\int_{0}^{t}{\bf { p}}\cdot {\bf {v}}(t^{\prime }){\rm
d}t^{\prime }]\left| \sigma \right\rangle .$ It follows that
$H_{V}(t)-{\bf {k}}(t)\cdot {\bf {J}}=-\bf {K}\cdot \bf {J}$ ,
where ${\bf {K}}$ is defined to be

\begin{equation}
 {\bf {K}}=k(\sin \lambda \cos \gamma ,\sin \lambda \sin
\gamma ,-\frac{1}{k}\dot{\gamma}(1-\cos \lambda )+\cos \lambda
),\eqnum{17}  \label{eq201}
\end{equation}
and consequently the module of ${\bf K}$ is $K=\varsigma k$ with

\begin{equation}
\varsigma =\sqrt{1-\frac{2\dot{\gamma}}{k}(1-\cos \lambda )\cos
\lambda +\left( \frac{\dot{\gamma}}{k}\right) ^{2}(1-\cos \lambda
)^{2}}.\eqnum{18}
\end{equation}
From the expression for $\left| \sigma ,{\bf {k}}(t)\right\rangle
$ in the interaction picture the helicity operator of the photon
is of the form $\frac{1}{K} \bf {K}\cdot \bf {J}$. Since we have
the time-evolution operator of wave function of photons, which
does not involve the chronological product, we can in principle
investigate the evolution and inversion of the helicity states of
photons propagating inside the curved fiber.

It is significant to apply the formulation presented here to the
Chiao-Wu case ( i. e., the case of photon moving along a helical
fiber ) in order to unfold the physical meanings of our results.
Consider the Chiao-Wu adiabatic cyclic case in which the rotating
frequency of photon moving on the helicoid reads
$\dot{\gamma}=\Omega $ with $\Omega =\frac{2\pi
c}{\sqrt{d^{2}+(4\pi a)^{2}}}$ where $d$ and $a$ respectively
denote the pitch length and the radius of the helix, and $c$ is
the speed of light. When the initial condition is taken to be
$k_{1}=0,k_{2}=0,k_{3}=k$, i. e., $\cos \lambda=1$, it is readily
verified that the initial expectation value

\begin{equation}
\left\langle \frac{1}{K}\bf {K}\cdot \bf {J}\right\rangle
=\frac{\sigma \lbrack \cos \lambda -\frac{\Omega }{k}(1-\cos
\lambda )]}{\sqrt{1-\frac{2\Omega }{k}(1-\cos \lambda )\cos
\lambda +\left( \frac{\Omega }{k}\right) ^{2}(1-\cos \lambda
)^{2}}}              \eqnum{19}    \label{eq6}
\end{equation}
of the helicity of photon is equal to $\sigma $. If, for example,
the length scales, $d$ and $a$, of the optical fiber are taken to
be $d\rightarrow 0,$ $a\rightarrow 0$, which means that the fiber
is curved extremely on a small enough diameter, and in consequence
the rotating frequency $\Omega $ of photon moving along the curved
fiber tends to $\infty $ ( and $\lambda $ is negligibly small ),
then the expectation value, $\left\langle \frac{1}{K}\bf {K}\cdot
\bf {J}\right\rangle $, of the helicity of photon in the evolution
process may approach $-\sigma $ with $\sigma =\pm 1$ corresponding
to the right- and left- polarized photon, respectively. The
detailed calculation is given as follows: since in this case the
rotating frequency $\Omega $ tends to be infinite and $\lambda $
is small, the fraction on the right-handed side of Eq. (\ref{eq6})

\begin{equation}
\left\langle \frac{1}{K}\bf {K}\cdot \bf
{J}\right\rangle\rightarrow \frac{\sigma \lbrack -\frac{\Omega
}{k}(\frac{\lambda ^{2}}{2})]}{\sqrt{(\frac{\Omega
}{k})^{2}(\frac{\lambda ^{2}}{2})^{2}}}=-\sigma.    \eqnum{20}
\end{equation}
Note that taking the Chiao-Wu case as an illustrative example here
does not mean that no attempt is made to achieve theoretical
rigor, since this case is a special limit ( i. e., an adiabatic
limit in which the optical fiber is helically curved ) in our
treatment of the propagation of photons moving in a noncoplanarly
curved Fiber. Hence, from what has been discussed above, we can
conclude without exaggeration that the formulation based on the
Lewis-Riesenfeld invariant theory, rather than on the Berry's
adiabatic phase theory, can satisfactorily deal with the helicity
inversion of photons in the arbitrarily curved optical fiber.

As is presented previously, in the case of curved fiber the
operator $\frac{1}{k}{\bf k}\cdot {\bf J}$ together with the
effective Hamiltonian $H_{eff}(t)$, can describe the motion of
photons in the fiber, for the reason that it is the conserved
operator with the {\it time-independent} eigenvalue $\sigma $, and
can therefore be considered a Lewis-Riesenfeld invariant. But here
it is worth noticing that in a curved optical fiber the helicity
of a propagating photon is defined to be $\frac{1}{K}{\bf K}\cdot
{\bf J},$ rather than the conventional definition $\frac{1}{k}{\bf
k}\cdot {\bf J}$ which has well described the state of photon in
free space. For the extremely curved fiber, the expectation value
of photon helicity may varies ( even inverses )\cite{Yuan} in the
propagation inside the fiber, and therefore the invariant
$\frac{1}{k}{\bf k}\cdot {\bf J}$ may no longer considered the
definition of photon helicity. It follows from (\ref{eq201}) that
in smoothly curved fiber ( i. e., the helix diameter is large
enough and consequently the angle $\lambda $ in the momentum space
is small ), or in the case where the rotating frequency,
$\dot{\gamma}$, vanishes, it is apparently seen that the operator
$\frac{1}{K}{\bf K}\cdot {\bf J}$  tends to the conventional
definition of photon helicity, $\frac{1}{k}{\bf k}\cdot {\bf J}$.

\section{Potential applications of helicity inversion to information science}
What interaction results in the helicity inversion of photons in
the curved optical fiber? It follows from both the equation of
motion (\ref{eq100}) and the time-evolution equation (\ref{eq1})
of photons that the generalized Lorentz force ( Coriolis force
)\cite{Shen3,Qi} may lead to the reversal of photon helicity. The
effective magnetic field $\frac{ \bf{{\bf k}}\times \dot{\bf{{\bf
k}}}}{k^{2}}$ causes the variations of the direction of ${\bf k}$
( the wave vector of photon ) and can therefore be coupled to the
generalized magnetic moment ( proportional to the total angular
momentum, ${\bf J}$, of photons ). It should be noted that the
interaction Hamiltonian ( effective Hamiltonian $H_{eff}(t)$ ) in
the present Letter is obtained via the infinitesimal rotation
operator of motion of photon in the fiber, rather than through
analyzing the electromagnetic interaction between the photons and
the medium of the optical fiber. This means that the effective
Hamiltonian is a phenomenological Hamiltonian, which is based on
the previous assumption that the wave vector $\bf k$ of photon is
always along the tangent of the fiber at each point at arbitrary
time\cite{Chiao}. This assumption enables us to treat the
propagation of photons in the fiber by using the purely geometric
method ( containing the infinitesimal rotation operator in
3-dimensional space ). Since Chiao-Wu theory was consistent with
the fiber experiment\cite{Tomita}, we think that our generalized
approach to this problem may also be appropriate to consider the
more general cases, e. g., the non-adiabatic non-cyclic cases.
Perhaps someone may argue that the investigation of photon-atom
interaction in the curved fiber based on the purely physical
mechanism is also essentially significant. This may be the case,
but from the point of view of us, when the theory is compared with
the experimental results, the previous assumption is still
necessary, so long as the length scale is larger than the
wavelength. Only in the extremely curved case ( in which the
length scales, $d\rightarrow 0$ and $a\rightarrow 0$, are less
than the wavelength ) should we take into account the detailed
physical mechanism on the basis of the electromagnetic interaction
between photons and media of the fiber. In the smoothly curved
case, however, the geometric description of the electromagnetic
interaction between them as presented in this Letter is
convenient. It is readily verified that the helicity inversion of
photons results from this interaction.

For the present, physicists' control over the behavior of photons
has spread to include the photon number, phase\cite{Duan} and
polarization\cite{Muller} of light wave. If we could engineer all
the degrees of freedom of photons, our technology would benefit.
Already, the fiber communication, which simply guides light, has
revolutionized the telecommunications industry. It is now possible
for physicists to carry the laser communication and quantum
computation into practice, with the development of the modern
optics, fiber technology, quantum information and so on. For
instance, in the last decades a new frontier ( photonic
crystal\cite{Oskar} ) has emerged with the goal to control the
behavior ( amplitude or photon number ) of light; Muller {\it et
al.}\cite{Muller} applied the polarization or spin of photons to
the optical communication, and Duan {\it et al.}\cite{Duan}
suggested the application of the adiabatic Berry's phase driven by
the laser field to quantum computation. Apparently, it is of
essential significance to control and utilize the degrees of
freedom of photons ( photon number, polarization, helicity,
geometric phase, etc. ) in information science and technology. It
is with these goals in mind that the helicity inversion of photons
is taken into account in the present Letter. In the curved optical
fiber, the interaction of the photon spin with the wave vector
causes the helicity inversion of the photon, which is in exact
analogy with the transition operation from $0$ to $1$ in digital
circuit and may be said to possess the potential applications in
information technology. This inversion of helicity is controllable
by manipulating the spatial shape and helix radius of the curved
optical fiber. Furthermore, since geometric phase contains the
global properties of the varying geometric shape of the fiber, and
the measurement of the time-dependence of geometric phase may
obtain the information about the varying geometric shape of the
fiber, we think that the possible application of geometric phase
of photons to the optical communication deserves considerations.

\section{Concluding remarks}
The present approach to the geometric phase, time evolution,
polarization as well as the helicity inversion of photons
propagating along the optical fiber has several advantages over
other treatments based on the classical Maxwell's theory and the
Berry's adiabatic quantum theory. In this Letter, we are concerned
with the non-adiabatic non-cyclic case. Furthermore, the adiabatic
cyclic geometric phase $2\pi \sigma (1-\cos \lambda )$ reduced
from (\ref{eq4}) is consistent with the result in Chiao-Wu theory,
and the Chiao-Wu case is therefore considered the adiabatic limit
of our case. Additionally, We show that when it is propagating in
the curved ( or noncoplanarly curved ) optical fiber, the helicity
of the photon varies with time, and the method in the present
Letter is appropriate to treat the time-dependence of helicity and
the helicity inversion. If the fiber is extremely curved, then the
helicity inversion will take place, i. e., the change of the
helicity of the photon is dependent strongly on the spatial
appearance and the helix radius of the noncoplanarly curved fiber.
Most of the treatments regarding the propagation of photons in the
fiber had no Hamiltonian describing the motion of photons in the
fiber, but the Hamiltonian is indeed very necessary in
investigating the time evolution of photons in the fiber, so, in
this Letter we construct an effective Hamiltonian and then
transform the problem of motion of photons in the optical fiber
into that of the time-dependent quantum spin
model\cite{Bouchiat,Mizrahi}. The invariant-related unitary
transformation formulation replaces eigenstates of the {\it
time-dependent} invariants with that of the {\it time-independent}
invariants through the unitary transformation\cite{Zhu,Maamache}
and thus obtain the explicit expression for the time-evolution
operator, instead of the formal solution associated with the
chronological product operator. It is apparently of essential
significance to obtain the exact solution of Eq. (\ref{eq1}),
since in the treatment of helicity inversion of photons in
arbitrarily noncoplanarly curved optical fiber, the exact
solution, rather than the adiabatic solution obtained based on the
Berry's phase theory, is of real importance. All these advantages
enable us to study the helicity inversion of photons in the fiber
in more detail. We hold that not only this research would be of
physical interest and essential significance both experimentally
and theoretically, but also the applications of both geometric
phase and helicity inversion of photons inside the fiber to
optical communication, quantum information and related fields
deserved further investigations.

Acknowledgements This project is supported in part by the National
Natural Science Foundation of China under the project No.
$90101024$. Great thanks are due to Xiao-Chun Gao who led us to
this subject and gave many helpful suggestions.

\end{document}